\documentstyle[aps]{revtex}
\makeatletter
\newif\ifpr@pstyle \pr@pstylefalse
\newif\ifnons@qeq  \nons@qeqfalse

        \newlength{\paperbaselineskip}
\newfont{\fourteencp}{cmcsc10 scaled\magstep2}
\newfont{\titlefont}{cmbx10 scaled\magstep2}
\newfont{\authorfont}{cmcsc10 scaled\magstep1}
\newfont{\fourteenmib}{cmmib10 scaled\magstep2}
        \skewchar\fourteenmib='177
\newfont{\elevenmib}{cmmib10 scaled\magstephalf}
        \skewchar\elevenmib='177
\newfont{\ninemib}{cmmib9} \skewchar\ninemib='177
\makeatletter
\newcommand\nonsequentialeqnum{
        \nons@qeqtrue
        \@addtoreset{equation}{section}
        \def\theequation{\arabic{section}.\arabic{equation}}}
\newif\ifp@bblock  \p@bblocktrue
\newcommand\nopubblock{\p@bblockfalse}
\newcommand\topspace{\hrule height 0pt depth 0pt \vskip}
\newcommand\p@bblock{\begingroup \tabskip=\hsize minus \hsize
        \baselineskip=1.5\ht\strutbox \topspace-2\baselineskip
        \halign to\hsize{\strut ##\hfil\tabskip=0pt\crcr
        \the\Pubnum\crcr\the\date\crcr}\vskip 10mm\endgroup}
\newcommand\YUKAWAmark{\hbox{
        \ifpr@pstyle\ninemib\else\elevenmib\fi
        Yukawa\hskip1mm Institute\hskip1mm Kyoto \hfill}}
\newtoks\date
\newtoks\Pubnum
\let\pubnum=\Pubnum
\Pubnum={}
\date={January 2000}
\newcommand{\frontpageskip}{\vspace{12pt plus .5fil minus 2pt}}
\def\@authoraddress{} \def\@title{}
\def\title#1{\gdef\@title{\frontpageskip
        \begin{center}{\titlefont #1}\end{center}\par}}
\def\@author#1{\frontpageskip\par\begin{center}{\authorfont #1}
        \end{center}
        \nobreak}
\def\author#1{\expandafter\def\expandafter\@authoraddress\expandafter
    {\@authoraddress{\@author{#1}}}}
\def\andauthor#1{\expandafter\def\expandafter\@authoraddress\expandafter
    {\@authoraddress{\frontpageskip\centerline{and}\@author{#1}}}}
\def\authors#1{\expandafter\def\expandafter\@authoraddress\expandafter
    {\@authoraddress{\frontpageskip\noindent #1}}}
\def\@address#1{\par\begin{center}{\sl #1}\end{center}\par}
\def\address#1{\expandafter\def\expandafter\@authoraddress\expandafter
    {\@authoraddress{\@address{#1}}}}
\def\andaddress#1{\expandafter\def\expandafter%
    \@authoraddress\expandafter
    {\@authoraddress{\par\centerline{\sl and}\@address{#1}}}}
\renewcommand{\thanks}[1]{\footnote{#1}}
\def\maketitle{\par
  \begingroup
       \def\thefootnote{\fnsymbol{footnote}}
        \thispagestyle{empty}
        \baselineskip=\paperbaselineskip
        \@maketitle
        \endgroup
        \setcounter{footnote}{0}
        \let\maketitle\relax \let\@maketitle\relax
        \let\@thanks\relax \let\@title\relax
        \let\@title\relax \let\@authoraddress\relax
        \let\thanks\relax}
\def\@maketitle{%
        \ifpr@pstyle\vspace{-1.0cm}\else\vspace{-1.7cm}\fi
        \YUKAWAmark\vskip0.6cm
        \ifp@bblock\p@bblock \else\hrule height 0pt \relax \fi
        \@title
        \@authoraddress
        }
\makeatother

\newcommand{\bra}[1]{\left\langle #1 \right|}
\newcommand{\ket}[1]{\left| #1 \right\rangle}
\newcommand{\Hhat}{\hat{H}}
\newcommand{\HhatMq}{\hat{H}_M(q)}
\newcommand{\hhat}{\hat{h}}
\newcommand{\hhatq}{\hat{h}(q)}
\newcommand{\hhatMq}{\hat{h}_M(q)}
\newcommand{\Fhat}{\hat{F}}
\newcommand{\Fhatd}{\hat{F}^\dagger}
\newcommand{\Fhatp}{\hat{F}^{(+)}}
\newcommand{\Fhatm}{\hat{F}^{(-)}}
\newcommand{\Hc}{{\cal H}}
\newcommand{\Nhat}{\hat{N}}
\newcommand{\Ghat}{\hat{G}}
\newcommand{\Qhat}{\hat{Q}}
\newcommand{\Rhat}{\hat{R}}
\newcommand{\Phat}{\hat{P}}
\newcommand{\That}{\hat{\Theta}}
\newcommand{\Pc}{\overcirc{P}}
\newcommand{\Qc}{\overcirc{Q}}
\newcommand{\Nc}{\overcirc{N}}
\newcommand{\Thc}{\overcirc{\Theta}}
\newcommand{\Fp}{F^{(+)}}
\newcommand{\Fm}{F^{(-)}}
\newcommand{\Rp}{R^{(+)}}
\newcommand{\Rm}{R^{(-)}}
\newcommand{\Ab}{\mbox{\boldmath $A$}}
\newcommand{\Bb}{\mbox{\boldmath $B$}}
\newcommand{\Db}{\mbox{\boldmath $D$}}
\newcommand{\Nb}{\mbox{\boldmath $N$}}
\newcommand{\Qb}{\mbox{\boldmath $Q$}}
\newcommand{\Pb}{\mbox{\boldmath $P$}}
\newcommand{\phit}{\phi(t)}
\newcommand{\phix}[1]{\phi(#1)}
\newcommand{\phiqppn}{\phi(q,p,\varphi,N)}
\newcommand{\phiqpn}{\phi(q,p,N)}
\newcommand{\qpn}{(q,p,n)}
\newcommand{\qpN}{(q,p,N)}
\newcommand{\phiq}{\phi(q)}
\newcommand{\Ts}{{\cal T}}
\newcommand{\del}{\partial}

\newcommand{\beq}{\begin{equation}}
\newcommand{\beqa}{\begin{eqnarray}}
\newcommand{\eeq}{\end{equation}}
\newcommand{\eeqa}{\end{eqnarray}}

\begin{document}

\pubnum{YITP-00-2}
\date{January 2000}

\title{Adiabatic Selfconsistent Collective Coordinate Method  
for \break 
Large Amplitude Collective Motion in Superconducting Nuclei}

\author{Masayuki Matsuo}
\address{
Yukawa Institute for Theoretical Physics, Kyoto University,
Kyoto 606-8502, Japan }

\author{Takashi Nakatsukasa}
\address{RI Beam Science Laboratory, RIKEN, Wako 351-0198, Japan}

\author{Kenichi Matsuyanagi}
\address{Department of Physics, Kyoto University, Kyoto 606-8502, Japan}


\maketitle

\vfill

\centerline{\fourteencp Abstract}

\vspace{5mm}
An adiabatic approximation to the selfconsistent collective
coordinate method is formulated in order to describe large
amplitude collective motions in superconducting nuclei
on the basis of the time-dependent Hartree-Fock-Bogoliubov 
equations of motion. 
The basic equations are presented in a 
local harmonic form which can be solved in a similar way as the
quasiparticle RPA equations.  The formalism guarantees 
the conservation of nucleon number expectation values. 
An extension to the multi-dimensional case is also discussed.

\vspace{30mm}

\section{Introduction}\label{sec:intro}

Large amplitude collective motions (LACM), such as 
fissions, shape transitions, anharmonic vibrations and
low energy heavy ion reactions, are often
encountered in the studies of nuclear structures and reactions. 
To go beyond the phenomenological models assuming some macroscopic
or collective degrees of freedom motivated by the experimental facts
and intuitions, many attempts have been made
to construct theories that are able to describe the LACM
on a microscopic basis of the
nuclear many-body Hamiltonian.
Especially, theories based on
the time-dependent Hartree-Fock (TDHF) approximation 
have been investigated extensively 
\cite{Villars,Baranger-Veneroni,Goeke-Reinhard,GRR,MP1,YK,MP,KWD,Rowe,KWDappl2,KWDappl,Rowe-Bassermann,Marumori,SCC,Matsuo,SCCappl}.
The TDHF is a general framework for describing low energy
nuclear dynamics accompanying evolution 
of the nuclear selfconsistent mean-field\cite{Ring-Schuck,Blaizot-Ripka}.
A LACM corresponds 
to a specific solution of the TDHF equation of motion.
Since such a solution forms only a subset of the whole TDHF states (Slater
determinants),  it is often called a collective path,  
a collective subspace, or a collective submanifold. 
The collective coordinates are then
a set of small number of variables that parameterize the collective
subspace, and
the collective Hamiltonian is a function governing the
time evolution of the collective coordinates. 
One of the main purposes of the LACM theories is to provide a 
scheme to determine 
the collective subspace  and the collective Hamiltonian
on the basis of microscopic many-body Hamiltonian.
Although the studies of LACM theories are the vast field of research 
with many recent developments to different directions,
realistic applications to nuclear structure problems are rather limited.
In this paper, we would like to propose
a new practical method to calculate the collective subspace.

The adiabatic approximation has been often utilized for formulating 
the theory of collective subspace.
Indeed, some class
of LACM, such as nuclear fissions, can be regarded as a slow motion,
thus justifying the adiabatic approximation. 
The adiabatic TDHF (ATDHF) theory
\cite{Villars,Baranger-Veneroni,Goeke-Reinhard}
is one of the most well known adiabatic theories and
has been applied in some cases to  realistic 
description of heavy ion reactions \cite{Goeke-Reinhard}.
The ATDHF theory, however, accompanied a problem of
non-uniqueness of the solution \cite{GRR,MP1}.
Later, efforts to 
settle the non-uniqueness problem were made from different view points.
The works of Ref.\cite{YK} emphasize the
importance of the canonical variable condition and the analyticity
as a function of collective coordinate for finding a unique
solution. The proposed procedure relying on the Taylor expansion
method has not been applied to realistic calculations.
Another work \cite{MP} pointed out that
the collective subspace can be uniquely determined
by using the next order equation of the ATDHF theory.
It is clarified also that the adiabatic collective path of
LACM becomes the valley line of the potential
function in the multi-dimensional space associated with the TDHF states
\cite{MP,KWD,Rowe}. Further, the adiabatic collective path can be defined
by equations for a local harmonic mode at
each point of the collective path. These developments are
summarized in a consistent way in the formalism of 
Ref. \cite{KWD}. 
Note however that the adiabatic theory of Ref. \cite{KWD} 
relys on a multi-dimensional classical phase space representation
of the TDHF determinantal states \cite{Ring-Schuck,Blaizot-Ripka}.
A realistic application of this theory has not been done yet
except for the one to a light nucleus \cite{KWDSi}. 
Furthermore a problem of particle number conservation arises when applied
to superconducting nuclei  \cite{KWDappl2}.

Theories without the adiabatic approximation have
been also developed within the TDHF framework.
The early works of this direction are called
local harmonic approximations 
\cite{Rowe-Bassermann,Marumori}. Later, 
a  set of general equations that can determine
the collective subspace and the
collective Hamiltonian
were found and formulated in a 
consistent form known as 
the selfconsistent collective coordinate method (SCC or SCCM) \cite{SCC}.
The theory is purely based on the TDHF with no further approximation.
The method  also provides a concrete and practical scheme 
to solve the basic equations
using a power series expansion with respect to the boson-like
variables defined as a linear combination of the collective
coordinates and momenta.
The pairing correlation in superconducting nuclei is easily
incorporated within the SCCM by adopting the time-dependent 
Hartree-Fock-Bogoliubov (TDHFB) equation
in place of the TDHF, and the conservation law of the particle number is
consistently introduced in the basic framework of the SCCM \cite{Matsuo}. 
Thanks to these features  the SCCM has been applied for many realistic
descriptions of anharmonic vibrations in medium and heavy nuclei
\cite{SCCappl}. However,
the expansion method may not be suitable for 
the large amplitude motions of
adiabatic nature, for which change of the nuclear mean-field
is so large that the power series expansion 
of the collective coordinates may not be justified.

In the present paper, we try to combine merits of two approaches
mentioned above, 
i.e. the SCCM and the adiabatic thoery in order to 
formulate a theory that provides a consistent and 
practical method easily applicable to realistic descriptions 
of the adiabatic LACM in superconducting nuclei.
We achieve this aim by
introducing an adiabatic approximation to the general framework
of SCCM. Here we treat superconducting nuclei since 
the pairing correlations play essential roles
in many cases, like spontaneous fission, tunneling between
superdeformed and normal deformed configurations, and
coupling between coexisting states with different nuclear shape
(shape coexistence phenomena).
Although the use of the superconducting mean field
requires us to respect the particle number 
conservation, the SCCM  allows a simple and consistent 
treatment of the conservation law. We also avoid the non-uniqueness
problem by utilizing the principles similar to that of Refs.
\cite{MP,KWD,Rowe}. Furthermore,
we shall  show that the
equations of the adiabatic SCCM thus formulated can be
transformed to another set of equations that have a similar structure
as the local harmonic approach to the adiabatic theories \cite{KWD}.  
Therefore, the present
formalism is not only an extension of the SCCM, but also
succeeds some aspects of
the recent adiabatic theories such as
Ref.\cite{KWD}. 

In addition to the general formulation (Sect.\ref{sec:form}), we
present a practical scheme to solve the basic equations 
given in the local harmonic form for general classes of the many-body 
nuclear Hamiltonian (Sect.\ref{sec:lha} and Appendix). 
These equations are given in terms of the matrix elements
of the many-body Hamiltonian expressed by the quasiparticle
operators, thus enabling ones to develop  a straightforward coding
of a numerical program to solve the equations. 
In this way, we provide a complete procedure to
extract the collective subspace and the collective Hamiltonian.
We also discuss a possible prescription to extend the formalism
to the cases of the multi-dimensional collective motions
(Sect.\ref{sec:multidim}).
Conclusions are outlined in Sect.\ref{sec:conclusion}.

\section{Basic Equations}\label{sec:form}

\subsection{The SCC method for superconducting nuclei}\label{sec:scc}

In this subsection, we recapitulate the basic equations of
the SCC method \cite{SCC} in a way suitable for treating the
superconducting nuclei.

We introduce the TDHFB approximation
to describe LACM in superconducting
many-fermion systems. Here the time-dependent 
many-body state vector $\ket{\phit}$ is 
constrained to a generalized Slater determinant, which is chosen 
as a variational wave function. Time evolution of
$\ket{\phit}$ is then determined by the time-dependent
variational principle

\begin{equation}\label{TDHFB}
\delta\bra{\phit}i{ \del \over \del t} - \Hhat\ket{\phit}=0,
\end{equation}
where the variation is given by 
$\delta\ket{\phit}=a^\dagger_\alpha a^\dagger_\beta\ket{\phit}$
in terms of the quasiparticle
operators $\{a^\dagger_\alpha, a_\alpha \}$ which satisfy the
vacuum condition $a_\alpha \ket{\phit}=0$.

We assume that the LACM can
be described in terms of the collective variables, i.e.
the collective coordinate and momentum $\{q,p\}$ that are 
variables parameterizing the TDHFB state vector.\footnote{
We focus our discussion on a case of single collective coordinate.
A multi-dimensional case is discussed in Sect.\ref{sec:multidim}.}
The whole space of the TDHFB state vectors can be parameterized
by $ M \times (M-1)$
 variables ($M$ being the number of the single
particle states) as shown by the generalized Thouless theorem 
\cite{Ring-Schuck,Blaizot-Ripka}.
A set of the TDHFB state vector $\ket{\phix{q,p}}$ 
forms the collective subspace in which the LACM can be properly described.
One of
the main problems we concern is how to determine the collective subspace
on the basis of the TDHFB equations of motion.
At the same time, we need to determine
the collective Hamiltonian $\Hc(q,p)$ that governs the equation of
motion for the collective variables $\{q,p\}$. 
This is a general purpose of theories of LACM.

When we apply the LACM theories to nuclei in the superconducting phase,
a special
attention has to be paid to the particle number conservation.
Since the TDHFB state vector is not an eigenstate of the particle number 
operator $\Nhat$,
one would like to formulate the LACM theory so that
the particle number expectation value is conserved during
the course of collective motion. This is a problem which is
specific to the TDHFB, but not to TDHF for which the state vector
is a number eigenstate.

It is well known \cite{Ring-Schuck} 
that the expectation value of a conserved observable
is kept constant during the time-evolution of $\ket{\phit}$
governed by the TDHF(B)
equations of motion.  In the case of the pairing problem,
the TDHFB state vector violates spontaneously
the symmetry with respect to the gauge rotation $e^{-i \varphi \Nhat}$,
but a rotational motion related to the gauge rotation 
(often called  the pairing rotation) emerges automatically
to restore the gauge symmetry.
Therefore, the LACM
of superconducting nuclei, described by the TDHFB theory,
necessarily accompany the pairing rotation,
for which we introduce the collective coordinate, $\varphi$, the gauge
angle, and the conjugate collective momenta, $N$, which 
represents the particle number \cite{Matsuo}. Thus, we are obliged to consider
a collective subspace that is parameterized by the set of
four collective variables $\{q,p, \varphi, N \}$.\footnote{
For simplicity, here we assume a single kind of particles.
Extension to systems with many kinds (e.g., protons and neutrons in nuclei)
is straightforward.}

Let us now present the basic equations of the SCCM that determine 
the collective subspace $\ket{\phi(q,p, \varphi, N)}$ and 
the collective Hamiltonian $\Hc(q,p,\varphi,N)$.
As discussed above, the variable $\varphi$ is introduced to represent
the gauge angle. This requirement is easily satisfied \cite{Matsuo}
if one uses the following parameterization 

\begin{equation}\label{rotating}
\ket{\phiqppn}=e^{-i \varphi \Nhat}\ket{\phiqpn},
\end{equation}
where $\Nhat$ is the number operator of particles. Here
$\ket{\phiqpn}$ represents an intrinsic state that rotates
in the gauge space.

Basic equations of the SCCM consists of
a canonical variable condition and
invariance principle of the time-dependent Schr\"{o}dinger equation
(TDHFB equation in our case).
The canonical variable condition is, in general, given by

\begin{mathletters}\label{cvc1}
\begin{eqnarray}
\label{cvc1a}
&&\bra{\phiqppn}i{\del\over\del q}\ket{\phiqppn}  =p + {\del S \over
\del q}, \\ 
\label{cvc1b}
&&\bra{\phiqppn}{\del\over i\del p}\ket{\phiqppn}  = - {\del S\over
\del p}, \\
\label{cvc1c}
&&\bra{\phiqppn}i{\del\over\del \varphi}\ket{\phiqppn} =N + {\del S\over
\del \varphi}, \\ 
\label{cvc1d}
&&\bra{\phiqppn}{\del\over i \del N}\ket{\phiqppn}  = - {\del S\over
\del N}, 
\end{eqnarray}
\end{mathletters}
for the collective subspace parameterized by two sets of coordinates 
$(q,\varphi)$ and momenta $(p,N)$. Although  $S$ is an arbitrary
function of $\{ q,p,\varphi,N\}$,
we choose $S=0$ 
which is appropriate for the adiabatic approximation\cite{YK}. 
Then the canonical variable
condition can be rewritten as equations for the state 
$\ket{\phi\qpN}$,

\begin{mathletters}\label{cvc2}
\begin{eqnarray}
\label{cvc2a}
\bra{\phiqpn}i{\del\over\del q}\ket{\phiqpn} & =p,  \\ 
\label{cvc2b}
\bra{\phiqpn}{\del\over i\del p}\ket{\phiqpn} & =0, \\ 
\label{cvc2c}
\bra{\phiqpn}\Nhat\ket{\phiqpn} & =N, \\ 
\label{cvc2d}
\bra{\phiqpn}{\del\over i \del N}\ket{\phiqpn} & =0. 
\end{eqnarray}
\end{mathletters}
The third equation requires that the collective variable $N$
is identical to the expectation value of the number operator.
In other words, the particle number expectation value does not
depend on the collective variables $(q,p)$ for the LACM under
consideration. This is nothing but the condition of particle number 
conservation.

The collective Hamiltonian is defined as value of the
total energy on the collective subspace, given by

\begin{mathletters}
\begin{eqnarray} 
{\cal H} & = \bra{\phiqppn}\Hhat\ket{\phiqppn} \\
         & = \bra{\phiqpn}\Hhat\ket{\phiqpn}.  \label{hcol}
\end{eqnarray}
\end{mathletters}
Since the Hamiltonian $\Hhat$ commutes with the number operator $\Nhat$,
the collective Hamiltonian does not depend on the gauge angle $\varphi$.
Therefore, $\varphi$ becomes cyclic as we expect.

The invariance principle of the TDHFB equation 
plays a central role to determine
the collective subspace, which requires
that the TDHFB state vector $\ket{\phi(q(t),p(t),\varphi(t),N(t))}$
evolving in time within the collective subspace
should obey the full TDHFB equation, Eq.(\ref{TDHFB}).
This is equivalent to a condition that the collective
subspace is an invariant subspace of the TDHFB equations of motion.
Inserting Eq.(\ref{rotating}) into the time-dependent variational
principle, Eq.(\ref{TDHFB}), one obtains

\begin{equation} \label{invpr}
\delta\bra{\phiqpn} \Hhat - {dq\over dt} \Pc + {dp \over dt} \Qc +
         {dN\over dt} \Thc - {d\varphi\over dt} \Nhat \ket{\phiqpn} 
                    = 0,
\end{equation}
where the infinitesimal generators defined by

\begin{mathletters}\label{gener}
\begin{eqnarray}
\Pc\ket{\phiqpn} & = i {\del \over \del q}\ket{\phiqpn}, \\
\Qc\ket{\phiqpn} & = {1\over i} {\del \over  \del p}\ket{\phiqpn}, \\
\Thc\ket{\phiqpn} & = {1\over i} {\del \over  \del N}\ket{\phiqpn}
\end{eqnarray}
\end{mathletters}
are used. These operators are one-body operators which can
be written as linear combinations of bilinear products 
$\{a^\dagger_\alpha a^\dagger_\beta, a_\beta  a_\alpha, 
a^\dagger_\alpha a_\beta\}$ of the quasiparticle operators 
defined with respect to $\ket{\phiqpn}$.
Because of the canonical variable conditions, these infinitesimal
generators satisfy the following commutation relations

\begin{mathletters}\label{comrel}
\begin{eqnarray} 
&\bra{\phiqpn}[ \Qc , \Pc ]\ket{\phiqpn} = i, \\
&\bra{\phiqpn}[ \Thc , \Nhat ]\ket{\phiqpn}  = i, 
\end{eqnarray}
\end{mathletters}
and commutators of other combinations of 
$\Qc,\Pc,\Thc,\Nhat$ give zero expectation value.
By taking the variation as 
$\delta\ket{\phiqpn}=\{\Pc,\Qc,\Thc,\Nhat\}\ket{\phiqpn}$,
Eq.(\ref{invpr})  produces 
the canonical equations of motion for the
collective variables

\begin{mathletters}\label{colmot}
\begin{eqnarray}
{dq \over dt} & = &{\del \Hc \over \del p} 
         = i \bra{\phiqpn} [ \Hhat , \Qc ] \ket{\phiqpn}, \\
{dp \over dt} & = & - {\del \Hc \over \del q} 
         = i \bra{\phiqpn} [ \Hhat , \Pc ] \ket{\phiqpn}, \\
{d\varphi \over dt} & = & {\del \Hc \over \del N} 
         = i \bra{\phiqpn} [ \Hhat , \Thc ] \ket{\phiqpn}, \\
{dN \over dt} & = & - {\del \Hc \over \del \varphi} = 0.
\end{eqnarray}
\end{mathletters}
Using Eq.(\ref{colmot}), Eq.(\ref{invpr}) then reduces to an
{\it equation of collective subspace}
\begin{equation}\label{eqcolsub}
\delta\bra{\phiqpn} \Hhat - {\del\Hc\over\del p} \Pc 
                          - {\del\Hc\over\del q} \Qc  
                          - {\del\Hc\over\del N} \Nhat \ket{\phiqpn} 
                    = 0.
\end{equation}
If we take a variation $\delta_\bot$ 
that is orthogonal to the infinitesimal generators 
$\{\Pc,\Qc,\Thc,\Nhat\}$, one can immediately show 
$ \delta_\bot \bra{\phiqpn} \Hhat\ket{\phiqpn} =0$, which
implies that the the energy expectation value is stationary on
the collective subspace for all the variations except for the
tangent directions along the collective subspace. In other words,
the collective mode is decoupled from the other
modes of excitation. 

We remark here that 
the above basic equations of the SCCM
are invariant under point transformations of the
collective coordinate
\begin{mathletters}\label{ptrans}
\begin{eqnarray}
&& q \rightarrow q'=q'(q), \\
&& p \rightarrow p'=p\times\left(\del q'/\del q\right)^{-1}.
\end{eqnarray}
\end{mathletters}
The basic principles, i.e. the
canonical variable condition, Eq.(\ref{cvc1}), and the invariance principle
of the TDHFB equation, Eq.(\ref{invpr}), are not affected by
the general canonical
transformations of collective variables 
$\{q,p,\varphi,N\} \rightarrow \{q',p',\varphi',N'\}$. 
By taking
the parameterization, Eq.(\ref{rotating}), 
and the specific choice of $S=0$ in Eq.(\ref{cvc1}),
the allowed canonical transformations are restricted to the point
transformations \cite{YK}.

\subsection{Adiabatic approximation} \label{sec:ascc}

Assuming that the LACM described by the collective 
variables $\{q,p\}$ is slow motion, we here introduce the
adiabatic approximation to the SCCM.
Namely we shall
expand the basic equations with respect to the collective
momentum $p$, which is appropriate for small value of momentum.
Since the particle number variable $N$
is a momentum variable in the present formulation,
we also expand the basic equations with respect to $n=N-N_0$,
when we consider a system with particle number $N_0$.

Let us first consider the expansion of the TDHFB state vector
$\ket{\phiqpn}$ in the collective subspace. The origin of the
expansion is the state $\ket{\phiq}\equiv\ket{\phiqpn}|_{p=0,N=N_0}$.
We can assume that this is a time-even
state, i.e., $\Ts \ket{\phiq} = \ket{\phiq}$ under the time-reversal
operation $\Ts$ (Here we consider the system of even numbers of particles).
Thanks to the generalized Thouless theorem, the state vector
$\ket{\phiqpn}$ is expressed as

\begin{equation} \label{thouless}
\ket{\phiqpn} = e^{i\Ghat\qpn}\ket{\phiq}
\end{equation}
by using the unitary transformation $e^{i\Ghat\qpn}$. Here the
hermitian operator $\Ghat$ is given by 

\begin{equation} \label{ig}
\Ghat\qpn = \sum_{\alpha > \beta} \left(G_{\alpha\beta}\qpn 
                 a^\dagger_\alpha a^\dagger_\beta
                  +G_{\alpha\beta}^*\qpn a_\beta a_\alpha \right)
           = \Ghat(q,p,n)^\dagger .
\end{equation}
Here and hereafter,
the quasiparticle operators 
$\{a^\dagger_\alpha, a_\alpha\}$ are always defined locally at each value
of $q$ and satisfy the condition $a_\alpha\ket{\phiq}=0$.
We now expand the operator $\Ghat\qpn$ in powers of $p$ 
and $n$
and keep only the lowest order term. Namely,

\begin{mathletters}\label{igexpand}
\begin{eqnarray}
\Ghat\qpn & = & p \Qhat(q) + n \That(q), \\
 \Qhat(q) & = & \sum_{\alpha > \beta}\left( Q_{\alpha\beta}(q)
                 a^\dagger_\alpha a^\dagger_\beta
                  + Q_{\alpha\beta}^*(q) a_\beta a_\alpha \right) 
              =\Qhat(q)^\dagger, \\
 \That(q) &= & \sum_{\alpha > \beta}\left(\Theta_{\alpha\beta}(q)
                 a^\dagger_\alpha a^\dagger_\beta
                  +\Theta_{\alpha\beta}^*(q) a_\beta a_\alpha \right)
             =\That(q)^\dagger.
\end{eqnarray}
\end{mathletters}
If we require that time-reversal  of $\ket{\phiqpn}$ causes
 sign inversion of the collective momentum $p$, i.e.
$\Ts \ket{\phiqpn} = \ket{\phi(q,-p,N)}$, the operators $\Qhat(q)$ and
$\That(q)$ are time-even ($\Ts\Qhat(q)\Ts^{-1}=\Qhat(q)$)
and time-odd ($\Ts\That(q)\Ts^{-1}=-\That(q)$), respectively. 
If we put $n=0$ (i.e. $N=N_0$), the parameterization
Eq.(\ref{thouless}) together with Eq.(\ref{igexpand}) reduces to 
$\ket{\phi(q,p)}=e^{ip\Qhat(q)}\ket{\phiq}$ , which is the same form 
as the one introduced by Villars 
 and often used in the ATDHF theories 
\cite{Villars,Goeke-Reinhard,MP}.

The collective Hamiltonian is expanded as

\begin{mathletters}\label{hcolexpand}
\begin{eqnarray}
 \Hc(q,p,N) & = & V(q) + {1\over 2} B(q) p^2 + \lambda(q)n, \\
 V(q) & = & \Hc\qpN|_{p=0,N=N_0} = \bra{\phiq}\Hhat\ket{\phiq}, \\
 B(q) & = &{1\over2}{\del^2\Hc\qpN\over\del p^2}|_{p=0,N=N_0} = 
-\bra{\phiq}[[\Hhat,\Qhat(q)],\Qhat(q)]\ket{\phiq}, \\
 \lambda(q) & = &{\del\Hc\qpN\over\del N}|_{p=0,N=N_0}
  = \bra{\phiq}[\Hhat,i\That(q)]\ket{\phiq}, 
\end{eqnarray}
\end{mathletters}
where we kept the collective momentum $p$ up to the second order,
while up to the first order in $n$. The collective Hamiltonian
for the system with $N=N_0$ particles ( $n=0$ ) is given by

\begin{equation}\label{hcolh2}
 \Hc(q,p,N_0) = V(q) + {1\over 2} B(q) p^2 
\end{equation}
as the sum of the collective potential $V(q)$ and the collective
kinetic energy (the second term).

We next expand the infinitesimal generators. It is convenient 
for this purpose to define the unitary transformation 
$\Pc' = e^{-i\Ghat}\Pc e^{i\Ghat}, \Qc' = e^{-i\Ghat}\Qc e^{i\Ghat}, 
\Thc' = e^{-i\Ghat}\Thc e^{i\Ghat} $ of the infinitesimal
generators $\Pc, \Qc, \Thc$. They are expanded  as

\begin{eqnarray}
\Pc' & = & \Phat(q) + e^{-i\Ghat}i{\del\over \del q}e^{i\Ghat} 
      =  \Phat(q) - p {\del \Qhat \over \del q} -n {\del \That \over
     \del q} + ... , \\
\Qc' & = & e^{-i\Ghat}{\del\over i \del p}e^{i\Ghat} 
      =  \Qhat(q) +{i\over 2}[\Qhat, p\Qhat+n\That] + ..., \\
\Thc' & = & e^{-i\Ghat}{\del\over i \del N}e^{i\Ghat} 
      =  \That(q) +{i\over 2}[\That, p\Qhat+n\That] + ..., 
\end{eqnarray}
with use of the general expansion formula
\begin{equation}
e^{-i\Ghat}\del e^{i\Ghat} = i \del \Ghat 
                        + {1 \over 2!} [i\del \Ghat, i\Ghat]
                        + {1 \over 3!} [[i\del \Ghat, i\Ghat],i\Ghat]
                        + ...
\end{equation}
The operator $\Phat(q)$ is the infinitesimal generator with
respect to $\ket{\phiq}$ defined by
\begin{equation} \label{phatq}
\Phat(q) \ket{\phiq} =  i {\del \over \del q}\ket{\phiq}.
\end{equation}
Similarly, we introduce the unitary transformation of the
number operator and expand it as

\begin{equation}
\Nc'  \equiv   e^{-i\Ghat}\Nhat e^{i\Ghat} 
      =  \Nhat + i [\Nhat, p\Qhat+n\That] + ... 
\end{equation}
Substituting these operators in
the canonical variable condition, Eq.(\ref{cvc2}), we have

\begin{mathletters}
\begin{eqnarray}
\bra{\phiq}\Pc'\qpN\ket{\phiq} &=& p, \\
\bra{\phiq}\Qc'\qpN\ket{\phiq} &=& 0, \\
\bra{\phiq}\Thc'\qpN\ket{\phiq} &=& 0, \\
\bra{\phiq}\Nc'\qpN\ket{\phiq} &=& N.
\end{eqnarray}
\end{mathletters}
Now we expand these equations with respect to momentum $p$ and $n$.

\underline{The zeroth order canonical variable conditions:}

\begin{eqnarray}
\label{cvcp}
&&\bra{\phiq}\Phat(q)\ket{\phiq} = 
   \bra{\phiq}i{\del\over \del q}\ket{\phiq}=0, \\ 
\label{cvcq}
&&\bra{\phiq}\Qhat(q)\ket{\phiq} = 0, \\ 
\label{cvct}
&&\bra{\phiq}\That(q)\ket{\phiq} = 0, \\ 
\label{ncons}
&&\bra{\phiq}\Nhat \ket{\phiq} = N_0.   
\end{eqnarray}
Eqs.(\ref{cvcq}) and ({\ref{cvct}) are automatically fulfilled by the
definition, Eq.(\ref{igexpand}),
of the operators $\Qhat(q), \That(q)$. Eq.(\ref{cvcp})
can be satisfied if the $q$-dependent phase of
$\ket{\phiq}$ is properly chosen. Eq.(\ref{ncons})
is nothing but the constraint on $\ket{\phiq}$ for
the conservation of average particle number. 

\underline{The first order canonical variable conditions:}

\begin{eqnarray}
\label{cvcdq}
&& \bra{\phiq}{\del \Qhat(q) \over \del q}\ket{\phiq} = -1, \\ 
\label{cvcqt}
&& \bra{\phiq}[\Qhat(q),\That(q)]\ket{\phiq} = 0, \\ 
\label{cvcqn}
&& \bra{\phiq}[\Qhat(q), \Nhat]\ket{\phiq} = 0.  
\end{eqnarray}
One finds 

\begin{equation}\label{cvcqp}
\bra{\phiq}[\Qhat(q),\Phat(q)]\ket{\phiq} = i ,
\end{equation}
which can be derived by differentiating Eq.(\ref{cvcq}) with respect to $q$
and using Eq.(\ref{cvcdq}).
One can also derive from Eq.(\ref{ncons})
\begin{equation} \label{cvcpn}
\bra{\phiq}[\Phat(q),\Nhat]\ket{\phiq} = 0 .
\end{equation}
These equations give constraints on the 
infinitesimal generators $\Qhat(q),\Phat(q)$ concerning 
the normalization, Eq.(\ref{cvcqp}), and the orthogonal condition to the
particle number operator, Eq.(\ref{cvcpn}).

Next we expand the equation of collective subspace, 
Eq.(\ref{eqcolsub}),
to obtain a complete set of the basic equations for the adiabatic
approximation. After rewriting Eq.(\ref{eqcolsub}) as

\begin{equation}
\delta\bra{\phiq}e^{-i\Ghat}\Hhat e^{i\Ghat} 
                          - {\del\Hc\over\del p} \Pc'
                          - {\del\Hc\over\del q} \Qc' 
                          - {\del\Hc\over\del N} \Nc'
     \ket{\phiq} = 0,
\end{equation}
one can expand each term with respect to $p$ and $n$ with use
of the equations listed above.

\underline{The zeroth order equation of collective subspace:}

\begin{equation} \label{eqcolsub0}
\delta\bra{\phiq}\Hhat  - \lambda(q)\Nhat - {\del V\over\del q} \Qhat(q) 
     \ket{\phiq} = 0.
\end{equation}

\underline{The first order equation of collective subspace:}

\begin{equation}\label{eqcolsub1}
\delta\bra{\phiq}[\Hhat  - \lambda(q)\Nhat, \Qhat(q) ] 
  - {1\over i} B(q) \Phat(q)
     \ket{\phiq} = 0.
\end{equation}

These equations are similar to the equations of path in the
Villars' ATDHF theory \cite{Villars} 
except that the present paper deals with the
superconducting Hartree-Fock-Bogoliubov (HFB) state, and that the
Hamiltonian accompanies the $q$-dependent chemical potential term 
$ - \lambda(q)\Nhat$. 
As we mentioned in Sect.\ref{sec:intro},
the ATDHF theory has the
problem that the solution satisfying these two equations
is not uniquely determined \cite{GRR,MP1}. Although 
an additional validity condition was introduced to further constrain the
solutions \cite{Goeke-Reinhard,GRR}, 
the procedure of Ref. \cite{Goeke-Reinhard} 
does not fully solve the problem since the method does not work around the
HF minima.

The non-uniqueness problem has been investigated
in recent developments of the adiabatic theories, and in our opinion
they are classified into two different approaches.
The first one represented by  Ref.\cite{YK} 
claims that the solution
is uniquely determined if
an RPA boundary condition is specified at the HF minimum
and if the analyticity of the collective path as a function of $q$
is imposed together with the canonical variable condition.
The solution, however, needs to be constructed in an analytic
way or by means of a Taylor expansion method with respect to the collective
coordinate $q$. We do not adopt this approach since we wish to construct
a method applicable to systems under large excursion from the HFB minimum.
We rather follow the other approach represented by Refs. 
\cite{MP,KWD,Rowe}. These theories require an additional condition 
that the equation of collective subspace 
(corresponding to the decoupling condition in Ref.\cite{KWD})
should be satisfied up to the next order of the adiabatic expansion.
In the present formulation,
this second order condition is expressed as follows.

\underline{The second order equation of collective subspace:}

\begin{equation} \label{eqcolsub2}
\delta\bra{\phiq}{1\over 2}[[\Hhat  - \lambda(q)\Nhat, \Qhat(q) ], \Qhat(q)] 
    -  B(q) \Delta\Qhat(q) 
     \ket{\phiq} = 0,
\end{equation}
where
\begin{eqnarray}
\label{DQ}
 \Delta\Qhat(q) = {\del \Qhat \over \del q} + \Gamma(q)\Qhat(q), \\ 
\label{Gamma}
 \Gamma(q)=- {1\over 2B(q)}{\del B\over \del q}.  
\end{eqnarray}
This equation is equivalent in its mathematical form
to the one given by Ref.\cite{MP} if 
the chemical potential term $-\lambda(q)\Nhat$ is neglected. 
The last term $-B(q)\Delta \Qhat(q)$, often called a curvature term,
was simply neglected in the original
version of local harmonic approximation \cite{Rowe-Bassermann,Marumori}.
In the next subsection, instead of neglecting this curvature term,
we shall rewrite $\Delta \Qhat(q)$
and change Eq.(\ref{eqcolsub2}) into a workable form.

It is worth noting
here the invariance of the adiabatic
equations against the coordinate transformation. The collective
momentum $p$ undergoes the linear homogeneous transformation under the
point transformation, Eq.(\ref{ptrans}). 
Therefore, different orders of the expansion
with respect to the power of $p$ are not mixed up under the 
point transformation. The invariance property of the basic
equations of SCCM  is thus inherited to each equation of the
adiabatic approximation listed above. One can also confirm
this property by seeing that the quantities appearing in 
the equations transform as

\begin{mathletters}
\begin{eqnarray}
&&\Qhat(q) \rightarrow \Qhat'(q') = \Qhat(q(q')) 
                    \left({\del q'\over \del q}\right), \\
&&\Phat(q) \rightarrow \Phat'(q') = \Phat(q(q')) 
                    \left({\del q'\over \del q}\right)^{-1}, \\
&&{\del V\over \del q} \rightarrow {\del V'\over \del q'} = 
      {\del V\over \del q}  \left({\del q'\over \del q}\right)^{-1}, \\
&&B \rightarrow B'(q') =  B(q(q'))
                    \left({\del q'\over \del q}\right)^{2},  \\
&&\Delta\Qhat(q) \rightarrow \Delta\Qhat'(q')=\Delta\Qhat(q). 
\end{eqnarray}
\end{mathletters}

\section{Local Harmonic Approximation to Collective Subspace} \label{sec:lha}

\subsection{Local Harmonic Equations} \label{sec:lhe}

In this section we give an approximate but concrete procedure to construct
a solution of the adiabatic SCC method.
To this end, we first derive, from the adiabatic equations,
another set of equations of collective subspace which can be
solved in a way similar to the RPA equation.

We first take a derivative of the zeroth order equation,
Eq.(\ref{eqcolsub0}), with respect to $q$, which leads to
\begin{eqnarray}
\label{eqcolsub2m}
&&\delta\bra{\phiq} [\Hhat -\lambda(q)\Nhat, {1\over i}\Phat(q)] 
 -C(q)\Qhat(q)
-{\del V \over \del q}\Delta\Qhat(q) -{\del \lambda \over \del q}\Nhat
 \ket{\phiq} =0, \\ 
&&C(q) = {\del^2 V \over \del q^2} - \Gamma(q){\del V \over \del q} ,
\end{eqnarray}
where $\Delta \Qhat(q)$ and $\Gamma(q)$ are given by
Eqs.(\ref{DQ}) and (\ref{Gamma}), respectively. 
Using Eqs.(\ref{eqcolsub2}),
we eliminate $\Delta \Qhat(q)$ and rewrite Eq.(\ref{eqcolsub2m}) as
\begin{equation}
\label{eqcolsub2n}
\delta\bra{\phiq} [\Hhat -\lambda(q)\Nhat, {1\over i}\Phat(q)]
 -C(q)\Qhat(q)
-{1 \over 2B(q)}
[[\Hhat-\lambda(q)\Nhat, {\del V \over \del q}\Qhat(q) ], \Qhat(q)]
-{\del \lambda \over \del q}\Nhat
 \ket{\phiq} =0 .
\end{equation}
Furthermore, due to Eq.(\ref{eqcolsub0}), we find
\begin{equation}
{\del V \over \del q}\Qhat=(\Hhat - \lambda\Nhat)_{A} ,
\end{equation}
where $(\Hhat - \lambda\Nhat)_{A}$ means
$a^\dagger a^\dagger$ and $aa$ part of the operator
$\Hhat - \lambda\Nhat$ containing two-quasiparticle creation and
annihilation in the normal-ordered expression.

We thus replace Eqs.(\ref{eqcolsub0})-(\ref{eqcolsub2}) by the equivalent set,
\begin{equation} 
\delta\bra{\phiq}\HhatMq \ket{\phiq} = 0,   \label{eqcsmf}
\end{equation}
\begin{equation} 
\delta\bra{\phiq}[\HhatMq, \Qhat(q) ] - {1\over i} B(q) \Phat(q)
     \ket{\phiq} = 0,     \label{eqcshq}
\end{equation}
\begin{equation}
\delta\bra{\phiq} [\HhatMq, {1\over i}\Phat(q)] -C(q)\Qhat(q)
-{1 \over 2B(q)}[[\HhatMq, (\Hhat - \lambda(q)\Nhat)_{A}], \Qhat(q)]
-{\del \lambda \over \del q}\Nhat
 \ket{\phiq} =0.  \label{eqcshp}
\end{equation}
In Eq.(\ref{eqcshq}) and (\ref{eqcshp}),
$\Hhat-\lambda\Nhat$ has been replaced by
\begin{equation} \label{Hmov}
\HhatMq = \Hhat - \lambda(q)\Nhat - {\del V\over\del q}\Qhat(q),
\end{equation}
since the last term has no influence. 
The operator $\HhatMq$ 
may be regarded as the Hamiltonian in the moving frame.
The second and third terms can be identified with generalized cranking terms
associated with the pairing rotation and the LACM,
respectively.

Equations (\ref{eqcshq}) and (\ref{eqcshp}) are linear equations
with respect to the one-body operators $\Qhat(q)$ and $\Phat(q)$.
They have essentially the same structure as
the standard RPA equations 
except for the last two terms in Eq.(\ref{eqcshp}).
The quantity $C(q)$ is the local
stiffness parameter defined as the second (covariant) derivative of the
collective potential $V(q)$. 
The infinitesimal
generators $\Qhat(q)$ and $\Phat(q)$ are thus closely related to
the harmonic normal modes locally defined for $\ket{\phiq}$ and
the moving frame Hamiltonian $\HhatMq$.
These equations may be called 
{\it local harmonic equations}. 

It was shown in Ref. \cite{MP} that
the zeroth, first and second order equations of ATDHF give
a valley line of a potential energy
surface in a multi-dimensional configuration space associated with 
the TDHF states.
Similarly,
the local harmonic equations we have obtained,
Eqs.(\ref{eqcsmf})-(\ref{eqcshp}),
define the valley of the multi-dimensional
potential energy surface. The solution of these
equations will be uniquely determined if a suitable boundary condition 
is specified. These features are similar to 
the formulation of Ref.\cite{KWD} 
where the valley equation of the potential energy surface is derived from
the decoupling condition.

We remark again that the local harmonic equations in the present
paper differ from
the ones of Rowe-Bassermann \cite{Rowe-Bassermann} and
Marumori \cite{Marumori} with respect to the the third and the
fourth terms of Eq.(\ref{eqcshp}), which arise from the curvature
term (derivative of the generator) and the particle number constraint,
respectively. It is important to keep the curvature term in order
to maintain the relation between the collective subspace and 
the valley of the potential surface. We also note that the present
formalism is invariant with respect to the point transformation
of the collective coordinate,
as is the formulation of Ref.\cite{KWD}.

\subsection{Matrix Formulation of Local Harmonic Equations} \label{sec:sollha}

Let us now give a procedure to find
the operators $\Qhat(q), \Phat(q)$ that satisfy 
the local harmonic equations, (\ref{eqcshq}) and (\ref{eqcshp}) ,
for a given state $\ket{\phiq}$.
Since they are linear equations with respect
to these operators, it can be solved in an analogous way to the standard RPA.
To show this, we first express the operator 
$\Phat(q)$ and $\Nhat$ in terms of the quasiparticle operators:

\begin{equation} \label{Phat}
\Phat(q) =  i \sum_{\alpha > \beta} \left( P_{\alpha\beta}(q)
                 a^\dagger_\alpha a^\dagger_\beta
                  - P_{\alpha\beta}^*(q) a_\beta a_\alpha \right)
          = \Phat(q)^\dagger, \\
\end{equation}
\begin{equation}
\Nhat =   \sum_{\alpha > \beta} \left( N_{\alpha\beta}(q)
                 a^\dagger_\alpha a^\dagger_\beta
                  + N_{\alpha\beta}^*(q) a_\beta a_\alpha \right).
\end{equation}
Note that the $a^\dagger a$ and c-number parts are neglected
here since they do not change the state vector $\ket{\phiq}$ except
for the phase. The Hamiltonian $\Hhat$ is also expressed in terms
of the same quasiparticle operators. Assuming that the matrix elements 
$Q_{\alpha\beta}, P_{\alpha\beta}$ are real, the local harmonic
equations are written as the following matrix equations.

\begin{mathletters} \label{lha}
\begin{eqnarray}
\label{lhap}
&&(\Ab-\Bb)\Qb -B(q)\Pb = 0, \\ 
\label{lhaq}
&&(\Ab+\Bb)\Pb - C(q)\Qb - {1 \over B(q)}\Db\Qb -\lambda'\Nb = 0,    \\ 
\label{orth}
&&\Pb^{T}\Nb=0, \\ 
\label{norm}
&&2\Qb^{T}\Pb = 1, \\
&& \lambda' = {\del \lambda \over \del q}. 
\end{eqnarray}
\end{mathletters}
Here all quantities are functions of $q$, and 
$\Qb= \left(...,Q_{\alpha\beta},...\right)^{T}$,
$\Pb= \left(...,P_{\alpha\beta},...\right)^{T}$, and
$\Nb= \left(...,N_{\alpha\beta},...\right)^{T}$,
are the vector representation of the matrix elements 
with  $\alpha>\beta$.  \Ab\ and \Bb\ are the matrices 
whose elements are given by
\begin{mathletters}\label{defAB}
\begin{eqnarray} 
(\Ab)_{\alpha\beta,\gamma\delta} &=& 
\delta_{\alpha\gamma}\delta_{\beta\delta}(e_\alpha+e_\beta) 
+ v^{22}_{\alpha\beta,\gamma\delta}, \\
(\Bb)_{\alpha\beta,\gamma\delta} &=& 
 v^{40}_{\alpha\beta\gamma\delta}, 
\end{eqnarray}
\end{mathletters}
in terms of the matrix elements of the moving frame Hamiltonian 
\begin{mathletters}\label{Hmovqp}
\begin{eqnarray} 
\label{Hmovqpmf}
\HhatMq &=& \sum_\alpha e_\alpha a^\dagger_\alpha a_\alpha \\ 
\label{Hmovqpx}
   && +  {1\over 4}\sum_{\alpha\beta\gamma\delta}
        v^{22}_{\alpha\beta,\gamma\delta}
       a^\dagger_\alpha  a^\dagger_\beta a_\delta a_\gamma \\ 
\label{Hmovqpv}
   && +  {1\over 4!}\sum_{\alpha\beta\gamma\delta}
        \left(v^{40}_{\alpha\beta\gamma\delta}
     a^\dagger_\alpha  a^\dagger_\beta a^\dagger_\gamma a^\dagger_\delta 
 + v^{04}_{\alpha\beta\gamma\delta}
       a_\delta a_\gamma a_\beta a_\alpha \right) \\ 
\label{Hmovqpy}
   && +  {1\over 3!}\sum_{\alpha\beta\gamma\delta}
        \left(v^{31}_{\alpha\beta\gamma,\delta}
     a^\dagger_\alpha  a^\dagger_\beta a^\dagger_\gamma a_\delta 
 + v^{13}_{\delta,\alpha\beta\gamma}
       a_\delta^\dagger a_\gamma a_\beta a_\alpha \right) .
\end{eqnarray}
\end{mathletters}
Here, due to Eq.(\ref{eqcolsub0}), $a^\dagger a^\dagger$ and $a a$ parts
of $\HhatMq$ vanish,
and $a^\dagger a$ part of $\HhatMq$ is diagonalized. The matrix
elements of the residual interactions in
Eqs.(\ref{Hmovqpx})-(\ref{Hmovqpy})
are antisymmetrized with respect to the quasiparticle indices.
The matrices \Ab\  and \Bb\ have the same 
structures as the ones often defined in the quasiparticle 
RPA formalism \cite{Ring-Schuck}. 
The matrix $\Db$ is defined by
\begin{equation} \label{defD}
(\Db)_{\alpha\beta,\gamma\delta} =
{1\over 2}\bra{\phiq}[[[\HhatMq,(\Hhat-\lambda(q)\Nhat)_{A}],
a^\dagger_\alpha a^\dagger_\beta + a_\beta a_\alpha],a_\gamma a_\delta]
\ket{\phiq}.
\end{equation}
These matrix elements are expressed also  
in terms of the Hamiltonian matrix elements as
\begin{mathletters} \label{Dmatel}
\begin{eqnarray}
 (\Db)_{\alpha\beta,\gamma\delta} &&=
 (d^{22}_{\alpha\beta,\gamma\delta}-d^{40}_{\alpha\beta\gamma\delta}
 +d^{11}_{\alpha\gamma}\delta_{\beta\delta}
 -d^{11}_{\beta\gamma}\delta_{\alpha\delta}
 -d^{11}_{\alpha\delta}\delta_{\beta\gamma}
 +d^{11}_{\beta\delta}\delta_{\alpha\gamma})/2, \\
d^{22}_{\alpha\beta,\gamma\delta} &&= \sum_\epsilon (
 v^{31}_{\alpha\beta\epsilon,\gamma}h_{\delta\epsilon} 
 -v^{31}_{\alpha\beta\epsilon,\delta}h_{\gamma\epsilon} 
 -v^{13}_{\alpha,\epsilon\gamma\delta}h_{\beta\epsilon} 
 +v^{13}_{\beta,\epsilon\gamma\delta}h_{\alpha\epsilon}), \\
d^{40}_{\alpha\beta\gamma\delta} &&= \sum_\epsilon (
 v^{31}_{\alpha\beta\gamma,\epsilon}h_{\epsilon\delta} 
 -v^{31}_{\beta\gamma\delta,\epsilon}h_{\epsilon\alpha} 
 +v^{31}_{\gamma\delta\alpha,\epsilon}h_{\epsilon\beta} 
 -v^{31}_{\delta\alpha\beta,\epsilon}h_{\epsilon\gamma}), \\
d^{11}_{\alpha\beta} &&= \sum_{\gamma > \delta} (
 v^{13}_{\alpha,\beta\gamma\delta}h_{\gamma\delta} 
 -v^{31}_{\gamma\delta\alpha,\beta}h_{\gamma\delta}), 
\end{eqnarray}
\end{mathletters}
where $h_{\alpha\beta}$ is the matrix elements of
$(\Hhat-\lambda\Nhat)_{A}$ defined by
\begin{equation} \label{defhmat}
(\Hhat-\lambda\Nhat)_{A} = \sum_{\alpha>\beta}h_{\alpha\beta}
(a^\dagger_\alpha a^\dagger_\beta + a_\beta a_\alpha).
\end{equation}
Note that \Db\ contains the matrix elements of the type 
$v^{13}$ and $v^{31}$.
These terms of the Hamiltonian do not contribute to the standard
RPA equations.

The solution of the matrix equations is obtained as follows.
 From Eq.(\ref{lha}), one obtains 
\begin{mathletters}
\begin{eqnarray}
\Qb &=& \lambda' B(q) \left((\Ab + \Bb)(\Ab - \Bb) - \Db - \Omega
\right)^{-1}\Nb, \\
\Pb &=& \lambda'
(\Ab - \Bb) 
\left((\Ab + \Bb)(\Ab - \Bb) - \Db - \Omega\right)^{-1}\Nb, 
\end{eqnarray}
\end{mathletters}
with 
\begin{equation} 
 \Omega = B(q)C(q). 
\end{equation}

The condition that the collective mode is orthogonal to the number operator,
Eq.(\ref{orth}),
gives the following equation
\begin{equation} \label{dispersion}
S(\Omega) \equiv  \Nb^{T} (\Ab - \Bb) 
   \left((\Ab + \Bb)(\Ab - \Bb) - \Db - \Omega\right)^{-1}\Nb  
   =0.
\end{equation}
The quantity $\Omega=B(q)C(q)$ represents the square of the 
frequency $\omega=\sqrt{BC}$ of the local harmonic
mode, which is not necessarily positive.
This equation can be regarded as a dispersion equation
to determine  $\Omega=\omega^2$ as a zero point of $S(\Omega)$. 
The normalization condition,
Eq.(\ref{norm}), then gives
a constraint on the value of $\lambda'^2 B(q)$.
The value of the mass parameter $B(q)$ is arbitrary, being related
to the invariance under the point transformation Eq.(\ref{ptrans}).
A choice of the coordinate, $q$, specifies a value of the mass parameter,
$B(q)$.
In practice, the coordinate is often scaled so as to make the mass
parameter unity.

When the residual interactions are the separable forces
such as the monopole pairing and the 
quadruple-quadrupole forces, the local harmonic equations reduce to
a simpler form. The dispersion equation
for the separable interaction does not require
a matrix inversion as in Eq.(\ref{dispersion}). 
The details are discussed in Appendix.

Ref.\cite{KWDappl2} has discussed a problem of spurious (Nambu-Goldstone)
modes for local harmonic approaches, and stated that the RPA equation
at non-equilibrium points must be extended in order to guarantee separation
of the spurious modes.
However, no practical way of solving the equation was given
because the equation has parameters which we do not have a method to calculate.
In our present formulation, the RPA equation is indeed extended to assure
the number conservation.

\subsection{Construction of Collective Subspace} \label{sec:subspace}

Let us finally give algorithms to construct the
collective subspace $\ket{\phiq}$ as a function of the collective
coordinate $q$. Note 
that the local harmonic equations,
Eqs.(\ref{eqcsmf})-(\ref{eqcshp}), are regarded as 
local equations in a sense that the equations can be solved
independently for different values of $q$. 
At the HFB ground state, $\ket{\phi_0}$, defined by the HFB equation
\begin{equation} \label{HFB}
\delta\bra{\phi_0} \Hhat -\lambda_0 \Nhat \ket{\phi_0} = 0 ,
\end{equation}
we find $\partial V/\partial q=0$.
Therefore, $\ket{\phi_0}$ is always a state
on the collective subspace because
Eq.(\ref{eqcsmf}) is automatically satisfied.
Eqs. (\ref{eqcshq}) and (\ref{eqcshp}) reduce to the standard
RPA equations at $\ket{\phi_0}$
since the last two terms in Eq.(\ref{eqcshp}) vanishes.
The operators $\Qhat,\Phat$ are then determined as  one of the normal modes
of the RPA equation.

For non-equilibrium states, in general,
Eq.(\ref{eqcsmf}) and
the other two equations, (\ref{eqcshq}) and (\ref{eqcshp}), are coupled.
We may solve the coupled equations in an iterative way.  As
discussed in Sect. \ref{sec:sollha}, 
one can find the operators $\Qhat(q)^{(n)}, \Phat(q)^{(n)}$
by solving  Eqs.(\ref{eqcshq}) and (\ref{eqcshp}) 
for a given trial state
$\ket{\phiq}^{(n)}$ ($n$ denoting the iteration step). 
This defines the moving frame Hamiltonian
$\HhatMq^{(n+1)} = \Hhat - \lambda(q)^{(n+1)}\Nhat - 
\left({\del V\over\del q}\right)^{(n)}\Qhat(q)^{(n)}$, which can
be used to construct a trial state $\ket{\phiq}^{(n+1)}$ for
the next iteration.
If the iteration converges, one obtains
a state $\ket{\phiq}$ on which Eqs. (\ref{eqcsmf})-(\ref{eqcshp}) are
simultaneously satisfied.
Repeating the same
procedure for different values of $q$, one finally obtains the 
collective subspace $\ket{\phiq}$ and the collective Hamiltonian
as a function of $q$. 

We remark here that
the operator $\Phat(q)$ thus determined does not guarantee Eq.(\ref{phatq}),
although the other equations are satisfied. In this sense,
the local harmonic solution is an approximate solution. 
The exact solution satisfying all the basic equations
in Sect.\ref{sec:ascc} may not exist in realistic situations. 
Only when the system is ``exactly decoupled'' \cite{KWD},
the above procedure gives the exact solution. 

It is possible to choose another algorithm which satisfies Eq.(\ref{phatq})
at the sacrifice of errors in Eq. (\ref{eqcsmf}).
Let $\ket{\phi(q_0)}$ be a solution
that satisfies the basic equations at $q=q_0$.
The infinitesimal generators $\Qhat(q_0), \Phat(q_0)$ 
are determined by
solving Eqs.(\ref{eqcshq}) and (\ref{eqcshp}).
Then one can generate the state $\ket{\phi(q_0 + \delta q)}$ 
for an infinitesimal shift of the collective coordinate as

\begin{equation}
 \ket{\phi(q_0 + \delta q)} = 
e^{-i \delta q \Phat(q_0)}\ket{\phi(q_0)} .
\end{equation}
Repeating this procedure, one can construct a collective subspace.
This solution should coincide with the one solved by the previous method
if the system is exactly decoupled. Difference between the
two gives a quality of decoupling for the collective subspace in
the adiabatic approximation.
The second procedure can be used to provide an 
initial guess, $\ket{\phiq}^{(0)}$,
for the iteration of the first method.

\section{Extension to multi-dimensional collective subspace}
\label{sec:multidim}

In this section we extend the adiabatic SCC method to
a case of multi-dimensional collective subspace
described by $D$ collective coordinates and conjugate
momenta $\{ q^i, p_i ; i=1, ..., D \}$. 

One can easily derive the basic equations of the adiabatic SCC method 
in parallel to the derivation given in Sections \ref{sec:form} 
by noting first that Eqs.(\ref{thouless},\ref{igexpand}) 
are now extended to 

\begin{eqnarray}
&& \ket{\phiqpn} = e^{i\Ghat\qpn}\ket{\phiq}, \\
&& \Ghat  =  p_i \Qhat^i(q) + n \That(q), 
\end{eqnarray}
where the operator $\Qhat^i(q)$ have now $D$ components having
coordinate label $i$.
It is implied here and hereafter 
that the same coordinate index ($i$ in the
above expression) appearing
in the super- and subscripts means to take the summation over it.
The infinitesimal generator $\Phat_i(q)$ have also $D$ components
each of which is related to the derivative 
$i{\del\over \del q^i}\ket{\phiq}$. In the following, the
coordinate dependence is often omitted. For instance, $B^{ij}(q)$ and
$\Qhat^i(q)$ will be simply denoted by $B^{ij}$ and $\Qhat^i$, respectively.

The adiabatic collective Hamiltonian is expressed as

\begin{eqnarray} \label{hcoln}
 \Hc(q,p,N) & = & V(q) + {1\over 2} B^{ij}(q) p_i p_j + \lambda(q)n .
\end{eqnarray}
The zeroth and the first order equations of the collective subspace
are derived as

\begin{eqnarray}
\label{eqcsn0}
&& \delta\bra{\phiq}\Hhat  - \lambda(q)\Nhat 
         - {\del V\over\del q^i} \Qhat^i 
     \ket{\phiq} = 0, \\ 
\label{eqcsnhq}
&& \delta\bra{\phiq}[\Hhat  - \lambda(q)\Nhat, \Qhat^i ] - {1\over i} 
B^{ij} \Phat_j 
     \ket{\phiq} = 0,  
\end{eqnarray}
while the second order equation becomes 

\begin{equation}
\delta\bra{\phiq}{1\over 2}[\Hhat  - \lambda(q)\Nhat, \Qhat^i , \Qhat^j] 
   + {1\over 6} [ {\del V\over \del q^k} \Qhat^k, \Qhat^i, \Qhat^j]
    -  {1\over 2}(B^{ik}\Qhat^j_{;k} + B^{jk}\Qhat^i_{;k}) 
     \ket{\phiq} = 0 \label{eqcsn2}
\end{equation}
with 
\begin{eqnarray}
\label{defQcdr}
&& \Qhat^i_{;j} = 
   {\del \Qhat^i\over \del q^j} + \Gamma^i_{kj}\Qhat^k,          \\ 
&& \Gamma^i_{kj} = {1\over 2} B^{il}\left(
                  {\del B_{lk} \over \del q^j}
                 +{\del B_{lj} \over \del q^k}
                 -{\del B_{kj} \over \del q^l} \right),
\end{eqnarray}
where $B_{ij}$ is the inverse matrix of $B^{ij}$
and the bracket including three operators defined by
\begin{equation} 
[A, B, C] = {1\over 2}([[A,B],C] +[[A,C],B]).
\end{equation}

Expanding the canonical variable condition with respect to $p_i$ and $n$,
the following equations are derived;

\begin{eqnarray}
&\bra{\phiq}\Phat_i \ket{\phiq} &= 0, \\
&\bra{\phiq}\Nhat \ket{\phiq} &= N_0 ,
\end{eqnarray}
and
\begin{eqnarray}
\bra{\phiq}[\Qhat^i,\Phat_j]\ket{\phiq} = i\delta_{ij}, \\
\bra{\phiq}[\Qhat^i, \Nhat]\ket{\phiq} = 0,  \\
\bra{\phiq}[\Phat_i,\Nhat]\ket{\phiq} = 0.
\end{eqnarray}

These basic equations are invariant
under the point transformation of the collective variables
\begin{mathletters}
\begin{eqnarray}
&& q^i \rightarrow q'^i=q'^i(q), \\
&& p_i \rightarrow p'_i=p_j\times\left(\del q^j/\del q'^i\right).
\end{eqnarray}
\end{mathletters}

We have adopted the vector-tensor notation \cite{Landau-Lifshiz} 
to manifest the transformation properties under the point
transformation. Quantities which have a coordinate index in 
the subscript and in the superscript have the transformation properties
of the {\it covariant} and {\it contravariant} vectors, respectively.
For example, 
\begin{eqnarray}
&& \Qhat^i \rightarrow \Qhat'^i = 
      \Qhat^j\times\left(\del q'^i/\del q^j\right), \\ 
&& \Phat_i \rightarrow \Phat'_i = 
      \Phat_j\times\left(\del q^j/\del q'^i\right).
\end{eqnarray}
The mass tensor $B^{ij}$ is the contravariant tensor of second rank.
The operator $\Qhat^i_{;j}$ defined by Eq.(\ref{defQcdr}) 
is the covariant derivative of
$\Qhat^i$, and $\Gamma^i_{kj}$ is the Christoffel symbol where the
mass tensor $B_{ij}$ plays the role of metric tensor. 

Let us now derive local harmonic equations of
collective subspace.  Taking the $q$-derivative,  the zeroth order
equation (\ref{eqcsn0}) leads to 

\begin{eqnarray}
\label{eqcsn0d}
&&\delta\bra{\phiq} [\Hhat -\lambda(q)\Nhat, {1\over i}\Phat_i] 
-C_{ij}(q)\Qhat^j
-{\del V \over \del q^j}\Qhat^j_{;i} -{\del \lambda \over \del q^i}\Nhat
 \ket{\phiq} =0, \\  
&&C_{ij}(q) = {\del^2 V \over \del q^i \del q^j} - 
\Gamma^k_{ij}{\del V \over \del q^k}. 
\end{eqnarray}
As we have done for the $D=1$ case,
we would like to eliminate
the covariant derivative $\Qhat^j_{;i}$ in Eq.(\ref{eqcsn0d}) 
in order to give a feasible form of the local harmonic equation.
This was done for the $D=1$ case
with help of the second order equation of the collective
subspace. The corresponding equations (\ref{eqcsn2}) 
give $D(D+1)/2$ 
constraints, while number of unknown parameters, $\Qhat^j_{;i}$, is $D^2$.
In fact, Eq.(\ref{eqcsn2}) is equivalent to

\begin{equation} \label{constQd}
\delta\bra{\phiq}{1\over 2}
[[\Hhat  - \lambda(q)\Nhat, \Qhat^j] ,\Qhat^i] 
   + {1\over 6} [[ {\del V\over \del q^k} \Qhat^k, \Qhat^j], \Qhat^i]
    -  (B^{ik}\Qhat^j_{;k} + \Rhat^{ij}) 
     \ket{\phiq} = 0, 
\end{equation} 
where $\Rhat^{ij}$ are arbitrary one-body operators which are
antisymmetric for exchange of indices $i$ and $j$. If we choose
$\Rhat^{ij}=0$, we can eliminate the derivative
term ${\del V \over \del q^j}\Qhat^j_{;i}$.
Then, Eq.(\ref{eqcsn0d}) leads to

\begin{equation} \label{eqcsnhp}
\delta\bra{\phiq} [\Hhat -\lambda(q)\Nhat, {1\over i}\Phat_i] 
-C_{ij}\Qhat^j
-{1\over 2}[[\Hhat  - \lambda(q)\Nhat, (\Hhat  - \lambda(q)\Nhat)_{A}],
B_{ij}\Qhat^j] -{\del \lambda \over \del q^i}\Nhat
 \ket{\phiq} =0. 
\end{equation}
This equation is an analog of Eq.(\ref{eqcshp})
 and is linear with respect to the
infinitesimal generators $\Qhat^i, \Phat_i$.
We can numerically solve Eqs.(\ref{eqcsn0}), (\ref{eqcsnhq})
and (\ref{eqcsnhp}) on 
the same line as in Sect.\ref{sec:sollha} and \ref{sec:subspace}.

It should be remarked that the local harmonic equation
Eq.(\ref{eqcsnhp}) for  $D>1$ are derived from
equations, (\ref{eqcsn0}) and (\ref{eqcsn2}),
but with an additional condition $\Rhat^{ij}=0$ in Eq.(\ref{constQd}).
This condition is introduced to obtain the local harmonic
equations parallel to the one-dimensional case.

\section{Conclusions}
\label{sec:conclusion}

We have formulated the adiabatic approximation to the general
framework of the selfconsistent collective coordinate method
in order to describe large amplitude collective motions in
superconducting nuclei. The formalism, based on the TDHFB equations
of motion, guarantees  the conservation of particle
number in a transparent way. 
We have shown that the equations of collective subspace 
are reduced to
local linear equations for the infinitesimal generators, which
can be solved with use of quasiparticle
representation of the Hamiltonian matrix elements. 
This provides a complete procedure
to determine the states $e^{ip\Qhat(q)}\ket{\phiq}$ 
in the collective subspace and 
the collective Hamiltonian $\Hc(q,p)= V(q) + {1\over 2}B(q)p^2$
as a function of the collective
coordinate $q$ and momentum $p$.
A possible extension to
the case of the multi-dimensional collective coordinates is also discussed.

We emphasize that 
the equations given in this paper are solvable by means of
the matrix method similar to the standard RPA.
We hope that
the present adiabatic theory is useful to solve number of open questions
in the realistic studies of nuclear large amplitude collective motion.

\appendix
\section*{Solution for the Separable Interactions}
\label{sec:app}

In this appendix, we give solutions of the local harmonic 
equations of collective subspace for the case where
the two-body interaction is given by the separable forces.
We assume that the Hamiltonian is given by
\begin{equation}
\Hhat = \hhat_0 - {\kappa\over 2}\Fhatd\Fhat,
\end{equation}
where $\hhat_0 (= \hhat_0^\dagger)$ and $\Fhat$ are one-body operators.
Equivalently, one may write 
\begin{eqnarray}
&& \Hhat = \hhat_0 - {\kappa\over 2}\Fhatp\Fhatp 
+ {\kappa\over 2}\Fhatm\Fhatm,  \\
&& \Fhat^{(\pm)} \equiv (\Fhat \pm \Fhatd)/2 = \pm \Fhat^{(\pm)\dagger}.
\end{eqnarray}

For the separable forces, it is customary to neglect the 
Fock term of the forces. This approximation is
easily and consistently implemented in the SCCM by assuming that
the equation of motion for the time-dependent mean-field state 
$\ket{\phi(t)}$ is now given by the time-dependent Hartree-Bogoliubov 
equation without the Fock terms,
\begin{eqnarray}
&& \delta\bra{\phit}i{ \del \over \del t} - \hhat(t)\ket{\phit}=0, \\
&& \hhat(t) = \hhat_0 
        - \kappa \Fhatp \bra{\phi(t)}\Fhatp\ket{\phi(t)}
        + \kappa \Fhatm \bra{\phi(t)}\Fhatm\ket{\phi(t)}.
\end{eqnarray}

The local harmonic equations (\ref{eqcsmf})-(\ref{eqcshp}) then become

\begin{equation}
\delta\bra{\phiq}\hhatMq \ket{\phiq} = 0,
\end{equation}
\begin{equation} \label{sephq}
\delta\bra{\phiq}[\hhatMq, \Qhat(q) ] - f^{(-)}_Q \Fhatm 
- {1\over i} B(q) \Phat(q) 
     \ket{\phiq} = 0,
\end{equation}
\begin{eqnarray} \label{sephp}
\delta\bra{\phiq} [\hhatMq, {1\over i}B(q)\Phat(q)] - f^{(+)}_P \Fhatp 
-B(q)C(q)\Qhat(q)
-f^{(+)}_R \Fhatp \nonumber \\
-[\Fhatm, (\hhat(q)-\lambda(q)\Nhat)_{A}] f^{(-)}_Q 
-f_N \Nhat
 \ket{\phiq} =0, 
\end{eqnarray}
where $\hhatMq$ is the mean-field Hamiltonian in the moving frame
defined by
\begin{eqnarray}
&& \hhatMq = \hhatq  - {\del V\over \del q}\Qhat(q)
                     -\lambda(q)\Nhat, \\
&&   \hhatq = \hhat_0  - \kappa \Fhatp \bra{\phiq}\Fhatp\ket{\phiq} ,
\end{eqnarray}
and definitions of other symbols are 
\begin{mathletters}
\begin{eqnarray}
&& f^{(-)}_Q = -\kappa \bra{\phiq}[\Fhatm, \Qhat(q)] \ket{\phiq}, \\
&& f^{(+)}_P = 
   \kappa \bra{\phiq}[\Fhatp,{1\over i}B(q)\Phat(q)] \ket{\phiq}, \\
&& f^{(+)}_R = 
   -\kappa \bra{\phiq}[[\Fhatp,(\hhatq -\lambda(q)\Nhat)_{A}],\Qhat(q)]
     \ket{\phiq}/2, \\
&& f_N = B(q){\del \lambda \over\del q}.
\end{eqnarray}
\end{mathletters}
We express all operators in the above equations in terms of the
quasiparticle operators $\{a^\dagger_\alpha, a_\alpha\}$
defined for $\hhatMq$ and $\ket{\phiq}$. For example, 
\begin{eqnarray}
&& \hhatMq =\sum_\alpha e_\alpha a^\dagger_\alpha a_\alpha, \\
\label{Fhatp}
&& \Fhatp = \sum_{\alpha>\beta} F^{(+)}_{\alpha\beta} 
(a^\dagger_\alpha a^\dagger_\beta +  a_\beta a_\alpha )
 + \sum_{\alpha\beta}F^{(+)}_{B,\alpha\beta} a^\dagger_\alpha a_\beta, 
   \\ 
\label{Fhatm}
&& \Fhatm = \sum_{\alpha>\beta} F^{(-)}_{\alpha\beta} 
(a^\dagger_\alpha a^\dagger_\beta -  a_\beta a_\alpha )
 + \sum_{\alpha\beta}F^{(-)}_{B,\alpha\beta} a^\dagger_\alpha a_\beta. 
\end{eqnarray}
We have assumed that all matrix elements are
real. Equations (\ref{sephq},\ref{sephp}) are then reduced to 
linear equations for the matrix elements
$Q_{\alpha\beta},P_{\alpha\beta}$ of the infinitesimal generators
$\Qhat(q),\Phat(q)$. They are easily solved to give the expression

\begin{eqnarray}
&&Q_{\alpha\beta} = 
 { e_\alpha + e_\beta \over (e_\alpha + e_\beta)^2 - \Omega} 
  F^{(-)}_{\alpha\beta} f^{(-)}_Q 
+ { 1 \over (e_\alpha + e_\beta)^2 - \Omega} 
\left( F^{(+)}_{\alpha\beta} f^{(+)}_{PR} 
       + R^{(-)}_{\alpha\beta} f^{(-)}_{Q}
       + N_{\alpha\beta} f_N \right), \\ 
&&B P_{\alpha\beta} = 
 { e_\alpha + e_\beta \over (e_\alpha + e_\beta)^2 - \Omega} 
\left( F^{(+)}_{\alpha\beta} f^{(+)}_{PR} 
       + R^{(-)}_{\alpha\beta} f^{(-)}_{Q}
       + N_{\alpha\beta} f_N \right) 
+ { \Omega \over (e_\alpha + e_\beta)^2 - \Omega} 
  F^{(-)}_{\alpha\beta} f^{(-)}_Q, \\
&& f^{(+)}_{PR} = f^{(+)}_{P}  + f^{(+)}_{R}, 
\end{eqnarray}
where we introduced the one-body operator
\begin{equation}
 \Rhat(q)^{(\pm)}  \equiv [ \Fhat_B^{(\pm)}(q), (\hhatq-\lambda(q)\Nhat)_{A}] 
                = \sum_{\alpha>\beta}
R^{(\pm)}_{\alpha\beta} 
(a^\dagger_\alpha a^\dagger_\beta \mp  a_\beta a_\alpha ),
\end{equation}
with $\Fhat_B^{(\pm)}(q)$ being the last terms of $\Fhat^{(\pm)}$ 
in Eqs.(\ref{Fhatp},\ref{Fhatm}). 

Inserting this expression for the definition of 
$f^{(+)}_{PR}, f^{(-)}_{Q}$, we obtain the equations for 
unknown quantities $f^{(+)}_{PR}, f^{(-)}_{Q}, f_N$. Similarly,
the condition of orthogonality to the number operator Eq.(\ref{orth}) 
gives another equation for $f^{(+)}_{PR}, f^{(-)}_{Q}, f_N$. 
They are summarized as a linear homogeneous equation which can be
written in a $3 \times 3$ matrix form

\begin{equation}
\left(
\begin{array}{ccc}
 & & \\
 & S_{x x'}(\Omega)& \\
 & & \\
\end{array}
\right) 
\left(
\begin{array}{c}
f^{(+)}_{PR} \\
f^{(-)}_{Q} \\
 f_N \\
\end{array}
\right)  = 0,
\end{equation}
where
\begin{mathletters}
\begin{eqnarray}
&& S_{11} = 2S^{(1)}_{\Fp\Fp} + S^{(2)}_{\Rp\Fp} - {1\over \kappa},    \\
&& S_{12} = 2\Omega S^{(2)}_{\Fp\Fm} + 2S^{(1)}_{\Fp\Rm} 
           + S^{(1)}_{\Rp\Fm} + S^{(2)}_{\Rp\Rm}, \\
&& S_{13} = 2S^{(1)}_{\Fp N} + S^{(2)}_{\Rp N},      \\
&& S_{21} = 2S^{(2)}_{\Fm\Fp}, \\
&& S_{22} = 2S^{(1)}_{\Fm\Fm} + 2S^{(2)}_{\Fm\Rm} - {1\over \kappa},    \\
&& S_{23} = 2S^{(2)}_{\Fm N}, \\
&& S_{31} = S^{(1)}_{N\Fp}, \\
&& S_{32} = \Omega S^{(2)}_{N\Fm} + S^{(1)}_{N\Rm}, \\
&& S_{33} = S^{(1)}_{NN}.
\end{eqnarray}
\end{mathletters}
The functions $S^{(1)}_{XY}$ with the symbols $X,Y$ denoting
$(X,Y) =  (\Fp,\Fp),$ $(\Fp,\Rm),$ $(\Fp, N),$ $(\Rp,\Fm),$ $(\Fm,\Fm),$ 
         $(N,N),$  $(N,\Fp),$  $(N,\Rm)$ 
are given by    
\begin{equation}
 S^{(1)}_{XY}  = \sum_{\alpha>\beta} 
 { e_\alpha + e_\beta \over (e_\alpha + e_\beta)^2 - \Omega} 
X_{\alpha\beta}Y_{\alpha\beta}, \\
\end{equation}
while the functions  $S^{(2)}_{XY}$ with
$ (X,Y) =  (\Fp,\Fm),$ $(\Rp,\Fp),$  $(\Rp,\Rm),$ $(\Rp, N),$ 
       $(\Fm,\Fp),$ $(\Fm,\Rm),$ $(\Fm, N),$ $(N,\Fm)$
are given by
\begin{equation}
 S^{(2)}_{XY} = \sum_{\alpha>\beta} 
 { 1 \over (e_\alpha + e_\beta)^2 - \Omega} 
X_{\alpha\beta}Y_{\alpha\beta}. \\
\end{equation}

The value of $\Omega$ is determined by finding the zero point
of  the dispersion equation 

\begin{equation}
{\rm det}\{S_{xx'}(\Omega)\} = 0.
\end{equation}
Normalizations of 
$f^{(+)}_{PR}, f^{(-)}_{Q}, f_N$ are fixed by the condition Eq.(\ref{norm}).
It is straightforward to extend the above procedure to the case where the
two-body interaction is given by a sum of the separable forces.


\end{document}